\documentclass{article}

\usepackage{PRIMEarxiv}

\PassOptionsToPackage{off}{epstopdf}
\usepackage[off]{epstopdf}

\usepackage{pgfplots}
\usepackage{pgfplotstable}
\usepackage[numbers]{natbib}
\pgfplotsset{compat=1.18}
\usepgfplotslibrary{groupplots}

\usepackage{hyperref}
\usepackage{xr-hyper}
\usepackage{xcolor}
\usepackage{soul}

\usepackage[T1]{fontenc}
\usepackage{inconsolata} 
\usepackage{pifont}

\newlength{\ttcharwd}
\AtBeginDocument{\settowidth{\ttcharwd}{\texttt{0}}}

\usepackage[utf8]{inputenc} 
\usepackage[T1]{fontenc}    
\usepackage{url}            
\usepackage{booktabs}       
\usepackage{amsfonts}       
\usepackage{nicefrac}       
\usepackage{microtype}      
\usepackage{lipsum}
\usepackage{fancyhdr}       
\usepackage{graphicx}       
\graphicspath{{media/}}     
\usepackage{verbatim}       

\usepackage{titlesec}
\titleformat{\paragraph}[runin]{\bfseries}{}{}{}[.]
\titlespacing*{\paragraph}{0pt}{0.7\baselineskip}{0.5em}

\usepackage{amsmath}
\usepackage{amsfonts}

\usepackage{tikz}
\usetikzlibrary{
  arrows.meta,
  positioning,
  calc,
  shadows.blur,
  fit,
  matrix,
  backgrounds
}
\usetikzlibrary{shapes.geometric, arrows, positioning}

\usepackage[british]{babel}
\usepackage{hhline}
\usepackage{multirow}
\usepackage{subcaption}
\usepackage{colortbl}
\usepackage{makecell} 
\usepackage{colortbl} 
\usepackage{makecell} 
\usepackage{amsmath} 
\usepackage{array}
\usepackage{caption}  

\usepackage{float}  

\usepackage{booktabs}
\usepackage{siunitx}
\usepackage{threeparttable}

\usepackage{listings}
\usepackage{xcolor}

\definecolor{pythonblue}{RGB}{32,74,135}
\definecolor{pythonstring}{RGB}{164,0,0}
\definecolor{pythoncomment}{RGB}{96,96,96}

\lstdefinestyle{pythonstyle}{
    language=Python,
    basicstyle=\ttfamily\small,
    commentstyle=\color{pythoncomment},
    stringstyle=\color{pythonstring},
    keywordstyle=\color{pythonblue},
    showstringspaces=false,
    numbers=left,
    numberstyle=\tiny\color{pythoncomment},
    frame=single,
    rulecolor=\color{black},
    backgroundcolor=\color{white},
    breaklines=true,
    breakatwhitespace=true,
    tabsize=4
}

\definecolor{backcolour}{rgb}{0.95,0.95,0.92}
\definecolor{commentcolour}{rgb}{0.5,0.5,0.5}
\definecolor{stringcolour}{rgb}{0.58,0,0.82}
\definecolor{keywordcolour}{rgb}{0,0,0.6}
\definecolor{numbercolour}{rgb}{0.1,0.1,0.5}
\definecolor{emerald}{RGB}{16,185,129}
\definecolor{crimson}{RGB}{220,38,127}
\definecolor{slate}{RGB}{100,116,139}

\title{Multi-Tool Robotics Enables \textit{In-Situ} Sample Manipulation for Time-Resolved Synchrotron Measurements}

\author{
  Aditya Bondada \\ 
  Center for Functional Nanomaterials\\
  Brookhaven National Laboratory \\
  Upton, NY 11973, USA\\
  \And
  Elizabeth M. Wall \\ 
  Department of Chemical and Biological Engineering\\ Princeton University\\ 
  Princeton, NJ 08540, USA
  \And
  Eric Yuan Xiao\\ 
  Department of Mechanical Engineering \\ Stony Brook University\\ 
   Stony Brook, NY 11790, USA  
 \And
  Quinn C. Burlingame \\  
  Department of Chemical and 
  Biological Engineering\\ Princeton University\\
  Princeton, NJ, 08540, USA
  \And
  Yueh-Lin Loo \\  
  Department of Chemical and 
  Biological Engineering\\ Princeton University\\
  Princeton, NJ, 08540, USA
  \And
  Esther H. R. Tsai\\
  Center for Functional Nanomaterials\\
  Brookhaven National Laboratory \\
  Upton, NY 11973, USA\\
  \texttt{etsai@bnl.gov} \\
  \And
  Ruipeng Li\\ 
  National Synchrotron Light Source II\\
  Brookhaven National Laboratory \\
  Upton, NY 11973, USA\\
  \texttt{rli@bnl.gov} \\
}

\begin{document}
\maketitle

\begin{abstract}
The high photon flux at synchrotron beamlines allows for the measurement of fast dynamical processes. However, beamline radiation-safety protocols prohibit human intervention during X-ray experiments, limiting the ability to perform versatile real-time sample manipulations during continuous data acquisition. 
Here we present a robotic platform at an X-ray scattering beamline to enable real-time sample handling and processing in the experimental hutch, revealing previously inaccessible transient \textit{in-situ} dynamics in perovskite thin films.
This modular multi-tool robotic architecture enables in-hutch sample manipulation beyond human-access constraints, establishing a foundation for automated and autonomous synchrotron experimentation.

\end{abstract}

\section{Introduction}

Synchrotron beamlines~\cite{mobilio2016synchrotron, willmott2019introduction, sedigh2020review, barbour2023x} provide exceptional temporal resolution due to their photon flux and fast detectors, enabling time-resolved measurements for \textit{in-situ} and \textit{operando} studies of material structural dynamics. However, despite these intrinsic capabilities, experimental workflows remain fundamentally constrained by safety and access-control procedures at synchrotron beamlines. 
In particular, the hutch that surrounds the beamline to protect users from stray radiation must be evacuated, searched, and sealed prior to starting X-ray operation. This introduces a \textasciitilde1 minute delay between manual sample manipulation and measurement acquisition.
This delay renders many transient processes, such as rapid phase transitions, experimentally inaccessible. In other words, it is impossible to capture critical intermediate states and nonequilibrium dynamics that emerge immediately following sample manipulation by a human operator.
Further time delays are also commonly introduced by the need to manually transport samples prepared elsewhere, such as a  chemistry laboratory, into the beamline before even beginning the hutch evacuation protocol. These labor-intensive, conventional experimental workflows are time-consuming and restrict access to transient information.

In recent years, robotics has evolved beyond traditional pick-and-place operations, with modern systems integrating computer vision, advanced sensing, and AI-driven control to enable complex multi-step workflows~\cite{mortazavi2025romu4o, halvorsen2025autonomous, shokry2025metarl, mao2024multimodal, gilles2024metagraspnetv2}. In chemistry and materials science, robotic systems are also enabling automation and closed-loop workflows~\cite{longley2026robinhood, wang2025autonomous, halder2025autobot, burger2020mobile}. 
These advances motivate the use of robotics for continuous, human-free operation in radiation-controlled synchrotron beamline environments to address key limitations of conventional experimental workflows.
%
Recent efforts have demonstrated the integration of robotic systems into synchrotron beamline workflows, highlighting their emerging role in experimental automation for sequential experimental operations such as solution processing, substrate handling, and other reproducible protocols.
Fernando et al.~\cite{fernando2024facile, fernando2025robotic_endstation} have demonstrated the integration of collaborative robots into the Bluesky data-acquisition framework at Brookhaven National Laboratory's National Synchrotron Light Source II (NSLS-II), addressing the mechanical, software, and safety challenges of robots in radiation environments. Ozgulbas et al.~\cite{zhang2023robotic_pendant} developed a robotic pendant-drop system for containerless X-ray photon correlation spectroscopy (XPCS) with microsecond resolution, illustrating how purpose-built robotic hardware can unlock measurement modalities impossible with manual operation. 
The Automated Macromolecular Crystallography (AMX) beamline at NSLS-II, an early adopter of beamline robotics as described by Schneider et al.~\cite{schneider2022amx, lazo2021robotic}, utilizes robotic sample changers and pipeline processing to acquire hundreds of datasets per day. Fearon et al.~\cite{fearon2024fragment_screening} shows how this throughput is transforming fragment-based drug discovery. 
Yamada et al.~\cite{yamada2022automated_totalscattering} extended full automation to X-ray total scattering at the Super Photon Ring - 8 GeV (SPring-8) beamline in japan, while 
Schaible et al.~\cite{schaible2025robotic_giwaxs} demonstrated high-throughput thin-film characterization at Lawrence Berkeley National Laboratory's Advanced Light Source (ALS) small angle X-ray scattering/wide angle X-ray scattering beamline.
Automation is also being integrated into the design of new beamlines~\cite{yazak2025nextiii}.

Halide perovskite thin films~\cite{zhao2022accelerated, liu2023highly, sidhik2024two, garai2026bypassing} have attracted broad interest for applications ranging from photovoltaics and light-emitting diodes to quantum materials due to their stoichiometrically-tunable optoelectronic properties and compatibility with low energy-input deposition methods such as slot-die coating, thermal evaporation, and spin-coating. 
The structure of solution-processed perovskite thin films is highly dynamic during deposition and thermal annealing and is sensitive to small changes in the process temperature, atmospheric conditions, and the content of the precursor solution. The final thermal annealing step is particularly critical, as heating drives the rapid conversion of the wet precursor film into a dry crystalline perovskite layer. Capturing these early structural dynamics is virtually impossible with manual experimentation at a beamline, as researchers must either place the wet film on a pre-heated stage prior to evacuating the hutch, and thus lose the first \textasciitilde1 min of dynamics data, or place the sample on a lower temperature stage and allow solvents to dry for \textasciitilde1 min before beginning collection while ramping up the heating stage temperature. The timescales and complexity of these these processes make perovskite fabrication a compelling testbed to integrate autonomous experimentation and robotic operation with \textit{in-situ} grazing-incidence wide-angle X-ray scattering (GIWAXS) experiments.\cite{AhmadiPerovskites, closeloopperovskites_} 

Here, we develop and deploy a multi-tool robotic platform at the NSLS-II 11-BM Complex Materials Scattering (CMS) beamline at Brookhaven National Laboratory.
The system integrates robotic manipulation, computer vision, and modular end-effectors, which we deploy in this work to dispense perovskite precursor solutions, spin-coat films, and thermally anneal them during continuous GIWAXS measurements. Robot-enabled execution of this multi-step workflow both reduces the overall processing time for \textit{in-situ} processing and also eliminates the missed observation windows inherent to manual workflows.   
This approach not only enables the collection of previously inaccessible structural evolution data immediately after sample manipulation, but also establishes a foundation for broader AI-driven automation at synchrotron facilities.

\section{Methods}

\subsection{Robotics Platform}

\begin{figure}
    \centering
    \includegraphics[width=0.86\textwidth]{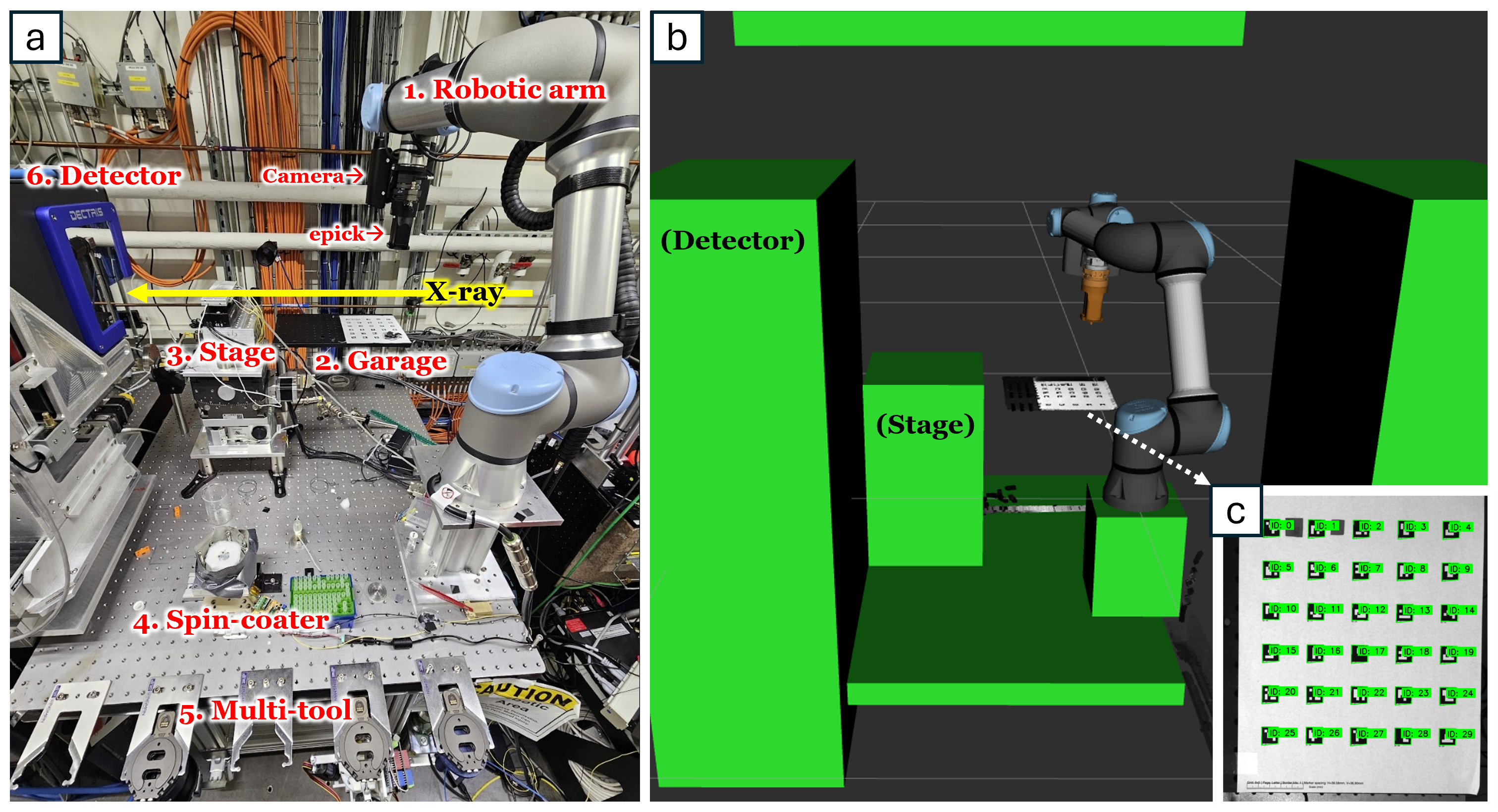}

    \caption{(a) Robotic platform on a modular table, including 1. robotic arm with camera, 2. sample garage, 3. annealing stage, 4. spin-coater and solutions for materials processing, 5. tool-exchange zone, and 6. Pilatus800k detector. The yellow line indicates the X-ray beam path. (b) RViz visualization of the planning scene with collision objects representing beamline hardware. The inset (c) shows the  live camera view and vision-based sample detection marked by ArUco tags in green. }
    \label{fig:cms_gui}
\end{figure}

A wide range of \textit{in-situ} experimental configurations can be accommodated on modular tables at the CMS beamline, allowing for quick switching between experiments and the deployment of specialized setups. Leveraging one such modular platform, we developed a platform for robotic sample manipulation.
As shown in Fig.~\ref{fig:cms_gui}(a), we deployed a modular robotic platform comprising (1) a UR5e collaborative robot arm (Universal Robots) integrated with (2) a barcode defined sample storage garage, (3) an annealing stage, (4) a solution processing station, and (5) a tool-exchange zone.
Figure~\ref{fig:cms_gui}(b) depicts the collision objects defined for safe robotic operation, with the inset Fig.~\ref{fig:cms_gui}(c) displaying a camera view from the robot.
Together, these components enable the robot to safely execute the complete experimental workflow from sample handling and preparation through \textit{in-situ} measurement without human intervention in the experimental hutch.
%

       
        
            
        

\subsubsection*{System Architecture}

The system is built on an ROS 2-based software architecture that provides a modular framework to coordinate robotic systems, motion stages, sample handling modules, and X-ray characterization through standardized communication and synchronized workflow control. 
Its node-based design enables flexible integration of additional hardware and experimental modules. Motion planning is handled through MoveIt 2, providing collision-aware trajectory generation that accounts for beamline hardware constraints and supports configurable exclusion zones for operation around sensitive beamline instrumentation. Visualization and monitoring are provided through RViz, as shown in Fig.~\ref{fig:cms_gui}(b), enabling real-time feedback of robot motion and environment mapping.

\begin{figure}
    \centering
    \includegraphics[width=0.85\textwidth]{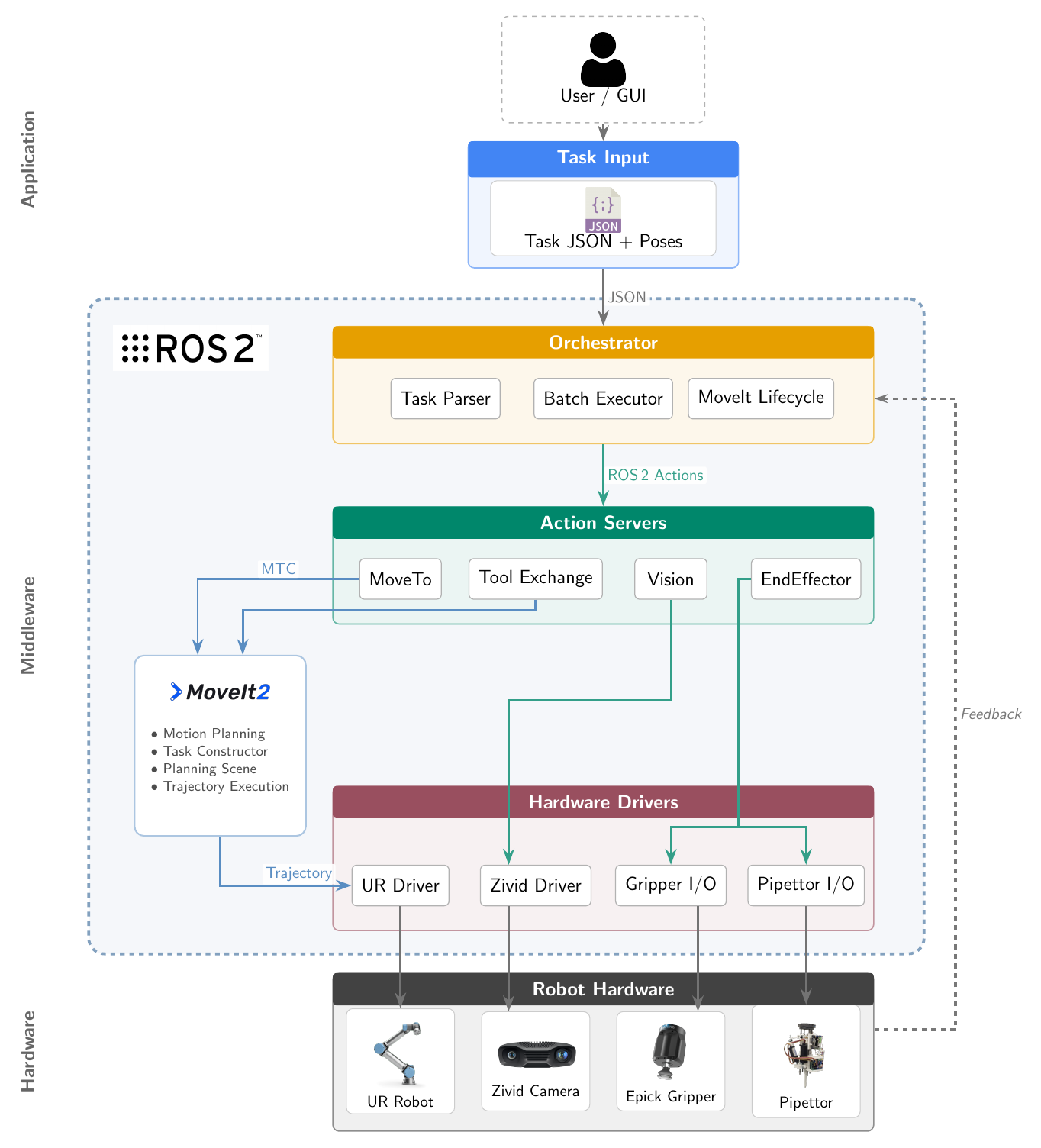}
    \caption{System architecture overview. Layered design spanning the application layer for experimental workflows and control, a middleware layer based on ROS 2 and MoveIt 2 for coordination and motion planning, and a hardware layer comprising robotic systems, motion stages, and beamline instrumentation.}
    \label{fig:architecture}
\end{figure}

As illustrated in Fig.~\ref{fig:architecture}, a three-layer software architecture comprising application, middleware, and hardware components is used in this work. 
At the application layer, scientists define experiments as structured JSON task files or through a graphical interface that provides task editing with live camera feedback, as shown in Fig.~\ref{fig:cms_gui}(c).
A ROS\,2 (Humble) middleware layer translates these task definitions into robot motions through MoveIt\,2, which generates collision-aware trajectories. Hardware drivers for the robot arm, end-effectors, and peripherals communicate over ROS\,2 topics and action servers.
Collision objects representing the beamline environment (sample stages, detector housings, and beam path exclusion zones) are registered in the MoveIt\,2 planning scene. All trajectories are planned against these constraints to avoid contact with beamline hardware.

\subsubsection*{Task Definition Framework}
\label{sec:task-framework}

Multi-step operations are programmed as an editable list of steps, allowing users to modify steps without changing the robot code.  Each operation is defined as a JSON task file containing a sequence of primitive operations, as shown in Fig.~\ref{fig:json-schema}. The file declares the starting end-effector, a dictionary of named joint poses, and an ordered list of primitives. Six primitive types are supported:

\begin{itemize}
    \item \texttt{vision\_scan}: moves the camera to predefined viewpoints, detects visible ArUco markers, averages their poses over repeated captures, and caches the results
    \item \texttt{moveto}: moves to a named joint configuration via a motion-planned trajectory
    \item \texttt{vision\_moveto}: moves to a cached marker pose obtained from a prior \texttt{vision\_scan}
    \item \texttt{end\_effector}: actuates the current tool (e.g., vacuum on/off, gripper open/close)
    \item \texttt{tool\_exchange}: docks the current tool and retrieves a specified replacement
    \item \texttt{pipettor}: actuates the custom pipetting end-effector for aspiration and dispensing; exposed as a separate primitive because the pipettor driver uses a different control interface
\end{itemize}

A typical experiment begins with a \texttt{vision\_scan} to localize all samples, followed by a sequence of motion and actuation primitives. Task files are version-controlled and can be shared across beamline visits to support reproducibility.


\begin{figure}[htbp]
\centering
\begin{tikzpicture}
\node[
    rectangle,
    rounded corners=3pt,
    draw=gray!40,
    fill=gray!5,
    inner sep=10pt,
    font=\ttfamily\footnotesize,
    align=left,
    text width=9cm,
] {
\textcolor{gray!60}{\{} \\
\quad \textcolor{teal!80!black}{"start\_gripper"}: \textcolor{orange!80!black}{"<gripper\_name>"}, \\
\quad \textcolor{teal!80!black}{"poses"}: \textcolor{gray!60}{\{} \\
\quad\quad \textcolor{teal!80!black}{"<pose\_name>"}: \textcolor{blue!70}{[j1, j2, j3, j4, j5, j6]}, \\
\quad\quad \textcolor{gray!50}{...} \\
\quad \textcolor{gray!60}{\}}, \\
\quad \textcolor{teal!80!black}{"tasks"}: \textcolor{gray!60}{[} \\
\quad\quad \textcolor{gray!60}{\{}\textcolor{teal!80!black}{"task\_type"}: \textcolor{orange!80!black}{"moveto"}, \textcolor{teal!80!black}{"target"}: \textcolor{orange!80!black}{"<pose>"}\textcolor{gray!60}{\}}, \\
\quad\quad \textcolor{gray!60}{\{}\textcolor{teal!80!black}{"task\_type"}: \textcolor{orange!80!black}{"vision\_moveto"}, \textcolor{teal!80!black}{"tag\_id"}: \textcolor{blue!70}{<id>}\textcolor{gray!60}{\}}, \\
\quad\quad \textcolor{gray!60}{\{}\textcolor{teal!80!black}{"task\_type"}: \textcolor{orange!80!black}{"end\_effector"}, \textcolor{teal!80!black}{"action"}: \textcolor{orange!80!black}{"<action>"}\textcolor{gray!60}{\}}, \\
\quad\quad \textcolor{gray!60}{\{}\textcolor{teal!80!black}{"task\_type"}: \textcolor{orange!80!black}{"tool\_exchange"}, \textcolor{teal!80!black}{"gripper"}: \textcolor{orange!80!black}{"<name>"}\textcolor{gray!60}{\}} \\
\quad \textcolor{gray!60}{]} \\
\textcolor{gray!60}{\}}
};
\end{tikzpicture}
\caption{Task definition schema. Each task file specifies the starting end-effector, named joint poses (6-DOF), and a sequence of task primitives. Supported task types include motion commands (\texttt{moveto}), vision-guided positioning (\texttt{vision\_moveto}), end-effector control, and tool exchange operations.}
\label{fig:json-schema}
\end{figure}

\subsubsection*{Multi-tool End-Effectors}
The UR5e supports end-effector exchange through a Wingman automatic tool changer (Triple A Robotics) mounted at the wrist, with tool docks located in the tool-exchange zone of the modular table. 
The vacuum gripper (Robotiq ePick) and custom pipettor enable the robot arm to perform both sample manipulation and liquid handling:
\begin{itemize}

\item A Robotiq ePick vacuum gripper transfers thin-film sample wafers between the sample garage and the thermal annealing stage. By utilizing vacuum suction instead of mechanical clamping, the system prevents structural damage to the delicate silicon or glass substrates; to prevent perturbations to the film surface when the sample is picked up, the point of contact is placed near the corner.
\item A custom pipettor end-effector was developed for this experiment, consisting of a linear actuator controlled by an Arduino microcontroller that operates a standard handheld pipettor. Interfaced via a serial connection to a ROS\,2 action server, this control stack enables precise aspiration and dispensing at specific coordinates orchestrated by the overarching task file. This custom design provides $\mu$L-scale dispense volumes with full control over actuation timing through the ROS\,2 task framework, suited to the small-volume liquid handling required for the experiments reported here. More broadly, the same integration pattern---a tool-changer-compatible wrist interface, a custom microcontroller, and a ROS\,2 action server---generalizes to other custom hardware developed for beamline experiments, allowing the platform to seamlessly hot-swap and incorporate tools beyond commercially available end-effectors.
\end{itemize}

\subsubsection*{Robotic Environment and Vision System}

Collision objects representing the beamline environment are defined as geometric primitives (boxes and cylinders) in a YAML configuration file, each specified by a reference frame, pose, and dimensions, then loaded into the MoveIt\,2 planning scene at startup. Obstacle bounds are determined from point-cloud captures acquired with the Zivid 2+ camera: each obstacle is fitted with a primitive sized to fully enclose its corresponding point cluster, providing a conservative safety margin. The beamline geometry is sufficiently static that this set of primitives provides repeatable collision-aware planning across deployments; updates to accommodate new instrumentation or relocated hardware are made by editing the YAML file directly.

Positions in the sample garage are recognized by a Zivid 2+ structured-light camera mounted on the wrist of the robotic arm. As shown in Fig.~\ref{fig:cms_gui}(c), a grid of ArUco markers is used to allocate thin film samples positioned next to known markers. During a \texttt{vision\_scan}, the camera moves to multiple predefined viewpoints and acquires several images at each. Marker poses are averaged across captures to reduce localization error. The averaged poses are cached, and subsequent \texttt{vision\_moveto} commands apply a fixed offset from the marker position to reach the corresponding sample.
Operations that do not require visual feedback, such as transferring samples to the measurement stage, changing tools, or dispensing solutions, use pre-taught joint configurations stored in the task file.

\subsection{Perovskite Film Preparation}
\label{sec:perovskite}

Formamidinium lead iodide (FAPbI$_3$) perovskite films were fabricated using a sequential deposition process. Si substrates were cleaned by sonicating in DI water, acetone, and isopropyl alcohol (IPA) for 5 minutes each. Immediately prior to solution deposition, the substrates underwent UV-ozone treatment for 15-20 minutes.
A 1.5M PbI$_2$ solution was prepared by dissolving PbI$_2$ in a mixture of dimethyl formamide (DMF) and dimethylsulfoxide (DMSO) with a 9:1 DMF:DMSO volume ratio. This solution was dispensed on a substrate and spin-cast at 1500 rpm for 30 s. The film was then annealed at 70$^{\circ}$C for 60 s to obtain a dry PbI$_2$ film.
A second solution of formamidinium iodide (FAI) and methylammonium chloride (MACl) was prepared by dissolving 90 mg FAI and 9 mg MACl in 1 mL IPA. This solution was dispensed dynamically atop the PbI$_2$ film while spin-coating at 2000 rpm. Spin-coating continued for 30 seconds after dispensing the solution. Finally, the film was transferred to a hotplate at 150$^{\circ}$C and annealed for 10 minutes to achieve the desired $\alpha$-FAPbI$_3$ phase. All of these dispensing, spin-coating, and annealing steps were performed via the multiple tools at the robotics platform at the beamline, with \textit{in-situ} GIWAXS collected during the final annealing step to quantify crystallization dynamics.

\section{Results}

The experiment is designed to replicate real experimental procedures previously performed in a chemistry lab~\cite{wall2025impact}, as described in Section~\ref{sec:perovskite}.
%
%
\textit{In-situ} GIWAXS measurements of the thin film samples were performed at an incident angle of 0.5$^{\circ}$ using a 200 $\mu$m (H) × 50 $\mu$m (V) X-ray beam with a wavelength of $\lambda$ = 0.9184 Å. 
A series of scattering patterns were collected in a rate of 10 patterns/s using a Pilatus 800K detector positioned 240mm downstream of the sample. The data analysis was carried out using SciAnalysis, ~\cite{SciAnalysis} a custom software package developed at NSLS-II.

\begin{figure}[H]  
    \centering
    \includegraphics[width=0.8\textwidth]{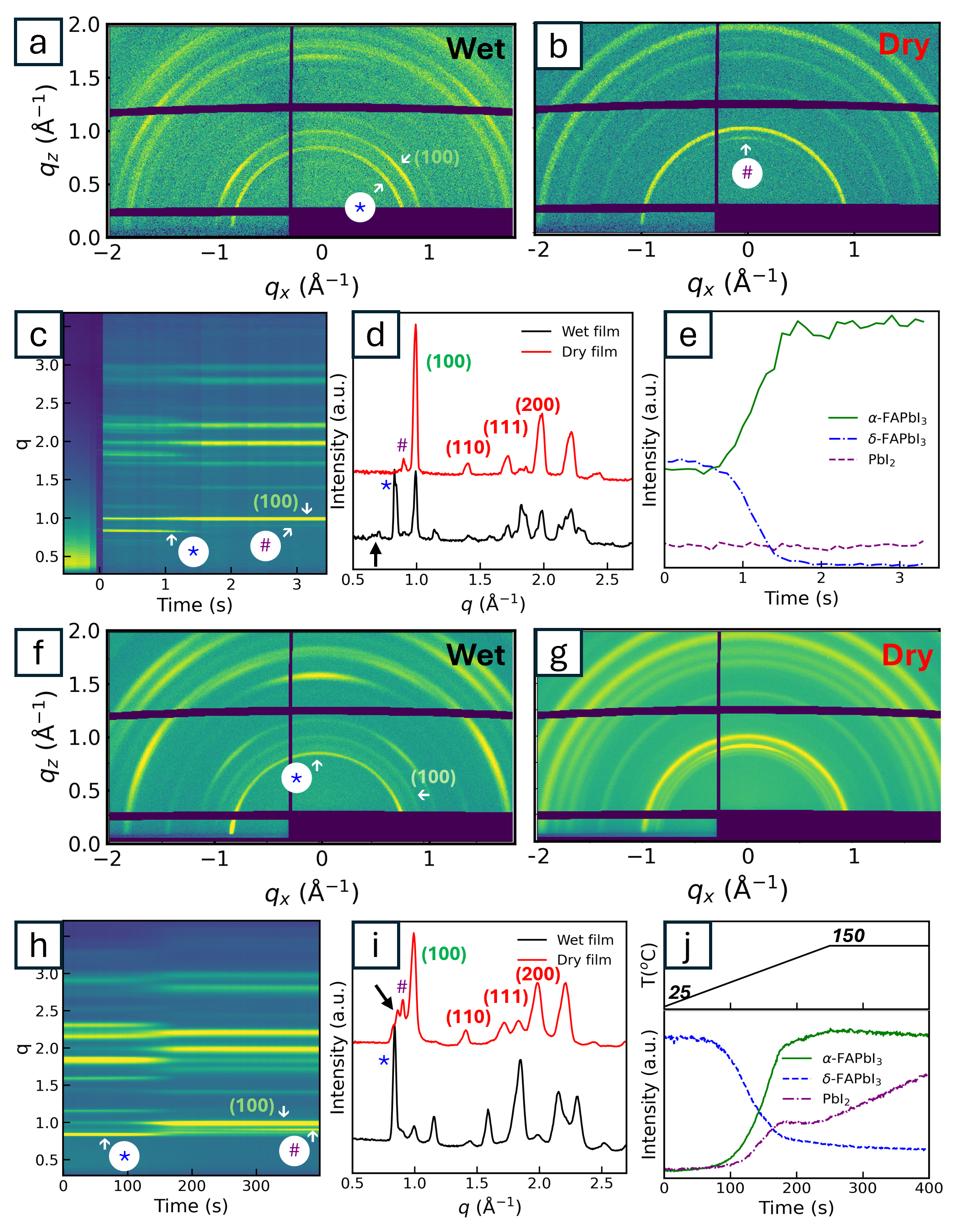}
    \caption{Comparison of thermal annealing process characterized by \textit{in-situ} GIWAXS on (a-e) robotic platform and (f-j) conventional thermal stage, including (a, f) scattering patterns at initial states of wet films and (b, g) final dry films, and (c-e, h-j) evolution of integrated intensities over time from wet to dry film. 
    In (j) are the temperature (top) and the intensities (bottom) vs time. The $\delta$-FAPbI$_3$ 2H (100) reflection is marked as blue *, the PbI$_2$ (100) in purple \#. 
    In (d, i), the indexing present the cubic phase of  $\alpha$-FAPbI$_3$, where the arrows mark the reflections of precursor-solvent complexes and the $\delta$-FAPbI$_3$ 6H (101), in (d) and (i) respectively.  
    }
    \label{fig:results}
    \hfill
\end{figure}

For experiments performed using the robotic platform, the custom-built hot stage was maintained at 150°C, with an integrated vacuum system used to hold the sample in a reproducible position after robotic delivery. After spin-coating and annealing the PbI$_2$ 1$^{st}$-step film, the FAI solution was spin-cast with the experimental hutch closed, and the robotic arm transferred the sample to the heating stage immediately after spin-coating. This rapid transfer of spin-cast samples is consistent with procedures for fabricating perovskite films in the laboratory. Data acquisition began prior to arrival of the the wet FAI-coated sample at the hot stage in order to capture the entirety of the structural evolution process. The scattering patterns for the initial (wet) and final (dry) states are shown in Figs.~\ref{fig:results}(a) and (b), respectively.
Time t=0~s represents the first frame in which the sample contacted the hot stage, as indicated in the diffraction plot in Fig.~\ref{fig:results}(c).
The initial state of the wet film consisted of a mix of precursor phases, including cubic $\alpha$-FAPbI$_3$ perovskite, hexagonal $\delta$-FAPbI$_3$ intermediates, PbI$_2$, and precursor-solvent complexes. The initial cubic $\alpha$-FAPbI$_3$ phase, identified by the (100) reflection at q=$1.0 \text{\AA}^{-1}$ showed preferred [111] orientation with respect to the substrate, with other indexed reflections shown in Fig.~\ref{fig:results}(d). The second most prominent phase had a primary reflection at q=$0.78 \text{\AA}^{-1}$, labeled with the purple $\#$ in Fig.~\ref{fig:results}, which we attribute to the (100) reflection of the hexagonal 2H phase of $\delta$-FAPbI$_3$. The presence of the $\delta$-phase in the wet film prior to annealing is consistent with previous \textit{in-situ} GIWAXS characterization of FAI spin-coating atop PbI$_2$ in an ambient environment with moderate (20-30\% RH) humidity levels~\cite{wall2025impact}. This hexagonal intermediate phase rapidly evolved into the cubic $\alpha$-FAPbI$_3$ perovskite phase within 1.5 s, evidenced by the concurrent decrease in (100) $\delta$-FAPbI$_3$ and increase in (100) $\alpha$-FAPbI$_3$ reflection intensities, shown in Fig.~\ref{fig:results}(c) and (e). Over the course of annealing, the oriented wet film transitioned to an isotropically-oriented dry film, as seen in Fig.~\ref{fig:results}(a) and (b). The weak reflection at q$\approx 0.70 \text{\AA}^{-1}$ likely corresponds to a FAI-PbI$_2$-DMSO complex~\cite{JACS_complexes}. Additionally, there is a feature from the (001) reflection of PbI$_2$ at q=$0.90 \text{\AA}^{-1}$ that does not change in intensity during the annealing process. It is common for sequentially deposited FAPbI$_3$ films to retain some residual PbI$_2$ at the conclusion of thermal annealing~\cite{wall2025impact}.


For comparison, \textit{in-situ} GIWAXS was also conducted using a manual thermal annealing procedure on a Linkam thermal stage at the beamline. The Linkam stage was kept at 25$^{\circ}$C while loading freshly spin-cast PbI$_2$ 1$^{st}$-step film. After the ~4 min it took to close the hutch and align the sample, \textit{in-situ} GIWAXS measurements were initiated at a rate of 1 pattern/s while the thermal stage was heated to 150$^{\circ}$C at a rate of 30$^{\circ}$C/min. Here, t=0 s corresponds to the onset of heating. The initial mixed phases at the start of the experiment still consisted of $\delta$-FAPbI$_3$,  $\alpha$-FAPbI$_3$, and PbI$_2$. However, the relative prevalence of these phases is quite different, and there was no longer evidence of precursor-solvent complexes, likely due to solvent evaporation during the several minute delay between spin-coating and the onset of the measurement (Fig. 4f). This manual experimental workflow, and the slow heating process in particular, also yielded a very different structural transition (Fig. 4g-j).  Evidence of phase transition was first observed at 90~s, corresponding to a stage temperature of 70$^{\circ}$C. From this time, the $\delta$-FAPbI$_3$ reflection intensity gradually decreased as the  $\alpha$-FAPbI$_3$ and PbI$_2$ reflections increased for approximately 100 s as temperature continued to ramp from 70$^{\circ}$C to 120$^{\circ}$C. The intensity of the PbI$_2$ reflection continued to increase even after the intensity of the FAPbI$_3$ reflections stabilized after 220~s. Additionally, we observed the emergence of a feature at q = $0.86 \text{\AA}^{-1}$ that we attribute to the (101) reflection of 6H $\delta$-FAPbI$_3$. In comparison with the robotic transfer, the manual process with slow heating resulted in a higher ratio of PbI$_2$ to $\alpha$-FAPbI$_3$ and incomplete conversion of the $\delta$-FAPbI$_3$ (Fig. 4i). The observed differences in the initial content of the films and in their dynamics over, particularly over the course of the slow heating experiments, indicate that the robot-assisted experimental process flow accesses a fundamentally different phase evolution pathway which is more representative of laboratory perovskite film fabrication.


\section{Discussion}


Sample and equipment handling in hard X-ray experimental hutches is limited by restricted access during beamtime, often constraining experiments to pre-programmed procedures. This limitation has become a bottleneck as demand grows for \textit{in-situ} and operando characterizations that require flexibility, rapid feedback, and iterative experimental control similar to modern materials research laboratories. 
Rapid heating can be achieved using techniques such as a photothermal annealing platform~\cite{leniart2020laser, yager2023autonomous_xray} or dedicated special thermal stages; however, implementing these methods typically requires the development and integration of such hardware, which is complex and problem-specific. 
In contrast, the multi-tool robotic platform provides a flexible and generalizable solution for sample manipulation in the experimental hutch, eliminating the need for dedicated instrumentation while replicating chemistry laboratory conditions and enabling a wide range of customized \textit{in-situ} experiments.

Beamline operation requires strict safety constraints to protect personnel and instrumentation. These are enforced through virtual boundaries within the ROS 2 framework, enabling safe integration of multiple tools while maintaining flexible operation. The system also supports advanced modules such as MoveIt 2 for motion planning and the Zivid ROS 2 driver with ArUco-based localization for vision-guided sample positioning. The platform achieves reliable accuracy and efficiency. 
In this work, ArUco fiducial markers were used primarily for sample stage localization, with future extensions incorporating additional components as needed.
Future developments will incorporate AI-driven machine vision to maintain safety operation with moving motors and detectors, and develop smart experimental planning, to optimize more sophisticated processes. 

While the UR5e manipulator has a manufacturer-specified pose repeatability of \SI{\pm 30}{\micro\meter}, the precision relevant to sample handling is set by the full integrated chain of the manipulator, tool flange, end-effector, vision pipeline, and sample-frame registration; In GI measurements, a beam size of 200 $\mu$m (H) × 50 $\mu$m (V) results in a footprint spanning several millimeters on the sample, reducing the demand for high-precision sample positioning.
Here the integrated system achieved approximately \SI{1}{\milli\meter} placement accuracy in vision-guided operations, which is sufficient in our GIWAXS studies, but may be consequential if sub-mm sample features are of interest. 
Refining camera-to-robot calibration and tool-specific offsets may further improve positioning accuracy.

The broader parameter space, including all processing conditions controlled at the beamline as well as sample transfer operations performed by the robotic system, defines a pathway toward establishing an operational laboratory environment directly at the beamline. This infrastructure can naturally support autonomous experimentation~\cite{yager2023autonomous_xray, doerk2023autonomous_morphologies, noack2021gaussian} in which real-time experimental data are used by decision-making algorithms to steer subsequent experimental actions. With robotics-enabled flexible in-hutch manipulation, the range of experiments that can be performed is significantly broadened, paving the way for an autonomous laboratory-at-the-beamline capable of supporting diverse processing workflows.


\section{Conclusion}

We have presented a multi-tool robotic platform for synchrotron scattering beamlines that enables multi-step processing and \textit{in-situ} characterization, bridging capabilities between laboratory workflows and beamline experiments. 
This approach allows us to capture the early-stage structural dynamics of perovskite crystallization beginning the moment the substrate is placed on a heated stage, which was previously inaccessible at synchrotron beamlines. 
Comparing the structural evolution data collected through fabricating samples with the robotic platform compared to those fabricated with a typical manual thermal annealing approach reveals distinct evolution pathways and final structures, which may help elucidate critical structure–property–performance relationships in materials.
The robotic platform provides a generalizable and modular solution for versatile \textit{in-situ} experiments. 
By eliminating the need for in-hutch human intervention, it enables measurements within previously inaccessible time windows and allows the observation of transient intermediate states and non-equilibrium dynamics.
Built on the ROS2 framework, the platform integrates motion control and machine vision to ensure safe, flexible, and extensible operation. Combining robotics-enabled manipulation with autonomous experimentation can enable diverse and autonomous laboratory-at-the-beamline experimentation.




\section*{Acknowledgments}
This research used beamline 11-BM (CMS) of the National Synchrotron Light Source II (NSLS-II) and utilized the X-ray scattering partner user program at the Center for Functional Nanomaterials (CFN), both of which are U.S. Department of Energy (DOE) Office of Science User Facilities operated for the DOE Office of Science by Brookhaven National Laboratory under Contract No. DE-SC0012704. 
This project was also supported in part by the U.S. Department of Energy, Office of Science, Office of Workforce Development for Teachers and Scientists (WDTS) under the Science Undergraduate Laboratory Internships Program (SULI).
YLL, QCB, and EMW acknowledge support from the U.S. Department of Energy's Office of Energy Efficiency and Renewable Energy (EERE) under Solar Energy Technologies Office (SETO) Agreement Number DE-EE0010503. 
We thank Dr. Phillip Maffettone for consulting the project.

\bibliographystyle{unsrtnat}  
\bibliography{ref_robot}


\end{document}